\documentclass[12pt]{article}
\usepackage{amssymb}

\input{tcilatex}

\begin{document}

\bigskip \baselineskip0.8cm \textwidth16.5 cm

\begin{center}
\textbf{First Order Phase Transition in the 3-dimensional Blume-Capel Model
on a Cellular Automaton }

N. Sefero\u{g}lu$^{\ast }$, A. \"{O}zkan$^{\ast }$ and B. Kutlu$^{\ast \ast
} $

$^{\ast }$ Gazi \"{U}niversitesi, Fen Bilimleri Enstit\"{u}s\"{u}, Fizik
Anabilim Dal\i , Ankara, Turkey

$^{\ast \ast }$ Gazi \"{U}niversitesi, Fen -Edebiyat Fak\"{u}ltesi, Fizik B%
\"{o}l\"{u}m\"{u}, 06500 Teknikokullar, Ankara, Turkey

e-mail: bkutlu@gazi.edu.tr
\end{center}

\textbf{Abstract}

The first order phase transition of the three-dimensional Blume Capel are
investigated using cooling algorithm which improved from Creutz Cellular
Automaton for the $D/J=2.9$ parameter value in the first order phase
transition region. The analysis of the data using the finite-size effect and
the histogram technique indicate that the magnetic susceptibility maxima and
the specific heat maxima increase with the system volume ($L^{d}$) at $%
D/J=2.9$.


\textbf{1 . Introduction}

The Hamiltonian of the Blume-Capel model $^{(1,2)}$ is given by,

\begin{equation}
H_{I}=-J\sum_{<ij>}S_{i}S_{j}+D\sum_{i}S_{i}^{2}
\end{equation}%
where $s_{i}=-1,0,1$ and the first sum is carried out over all
nearest-neighboring (nn) spin pairs on a three-dimensional simple cubic
lattice. The parameters of the $J$ and $D$ are the bilinear interaction
energy and single-ion anisotropy constant, respectively.

The model is known to have a rich critical behavior and considered by both
numerical and analytical methods. Recently, most of these studies mainly
focused on determining the tricritical point of the model$^{(3-8)}$. In
previous paper$^{(9)}$, we obtained the tricritical point value of the 3-d
Blume-Capel model as $D/J=2.82$. This results agrees with a series expansion 
$^{(3-4)}$, the cluster variation method $^{(5)}$, the effective field
theory $^{(6)}$ and Bethe- Peierls approximation results$^{(7)}$. Although
the critical behavior on the second-order phase transition region ($D/J<2.82$%
) has been studied commonly, the finite-size effects in the first-order
phase transition region ($D/J>2.82$) is not investigated exactly.

Our aim of this paper is to identify the phase transition in the first order
phase transition region of the Blume Capel model by the finite-size effects
and the histograms of energy distribution P(E) and order parameter
distribution P(M).

It is well known that the first order phase transitions involve the
coexistence of two distinct phases and are characterized by the existence of
a discontinuity in the energy and magnetization for the infinite systems. As
a results of these discontinuities, the specific heat and the susceptibility
show singularities of the delta function. However, the characteristic
singularities and discontinuities in the first order phase transitions
appears as rounded and smeared in finite systems employed in computer
simulations$^{(10,11)}$. These behaviors at the first order phase
transitions in finite-size systems are qualitatively similar to the
finite-size effects at second-order phase transitions. The finite size
effects at the first order phase transition have been investigated by the
different methods such as the Monte Carlo$^{(12-15)}$, the phenomenological
renormalization group$^{(16)}$, transfer matrix method$^{(17)}$,
renormalization-group analysis$^{(18)}$. According to the results of these
studies, the finite-size effects at the first order phase transition depend
on the volume of the system $L^{d}$. The specific heat and the
susceptibility increase with $L^{d}$ and the transition temperature which
are locations of their extrema, approach the transition temperature in the
infinite system as $L^{-d}$. In addition, the histograms of energy
distribution P(E) and the order parameter distribution P(M) are the reliable
method to study first-order phase transitions$^{(10,12,13,19)}$. While the
probability distribution of energy shows a single peak in a second-order
phase transition, it shows a double peak in a first-order phase transition.
In this paper we simulate the three dimensional Blume-Capel model with the
cooling algorithm which improve from the Creutz Cellular Automaton in the
first order phase transition region. The finite-size scaling and the
probability distribution for energy and order parameter are used to
determine the nature of the phase transition. The remainder of the paper is
organized as follows. The data are analyzed and the results are discussed in
Section 2 and the conclusion is given in Section 3.

\textbf{3. Results and discussion}

The three dimensional Blume-Capel model is simulated with the cooling
algorithm$^{(9)}$ which improved from the Creutz Cellular Automaton. The
simulations have been made at the $D/J=2.9$ anisotropy parameter value in
the first order phase transition region on simple cubic lattices $L$x$L$x$L$
of the linear dimensions $L=8,10,12,14,16,18$ and $20$ with periodic
boundary conditions.

At the algorithm, the cooling rate is equal to $0.01H_{k}$ per site for $%
D/J=29/10$ value and \ the kinetic energy of the system reduced by the
different cooling amounts per site because the kinetic energy, $H_{k}$, is
an integer variable in the interval $(0,24J)$. The computed values of the
quantities are averages over the lattice and over the number of time steps $%
(1.000.000)$ with discard of the first $100.000$ time steps during which the
cellular automaton develops.

The simulations were done about $10$ times with different initial
configurations at each lattice size. Measurements shown that the variation
of thermodynamic quantities and the histogram of P(E) and P(M) for each
different initial configuration are different from each other around the
transition temperature. At the same time, some of the simulations do not
exhibit the double peak in the histogram of the P(E) and the three peaks in
the histogram of the P(M) which characterize the first order phase
transition. Therefore, we choose the simulations which exhibit the
characteristic double peak in the histogram of P(E) and the three peaks in
the histogram of the P(M) to study finite-size effects at $D/J=2.9$.

In Fig.1, the temperature variation of the order parameter ($M=\frac{1}{L^{3}%
}\sum_{<ij>}S_{i}$ and $Q=\frac{1}{L^{3}}\sum_{<ij>}S_{i}^{2}$), the energy (%
$E=-J\sum_{<ij>}S_{i}S_{j}/E_{0}$ where $E_{0}$ is the ground state Ising
energy at kT/J=0), and histograms of $P(M)$ and $P(E)$ are shown for a
chosen simulation result on $L=18$ as an example. There are discontinuities
which characterize a first order phase transition in temperature dependence
of the order parameter and energy in Fig.1 (a) and (c). On the other hand,
another evidence for the first order transition is seen in the histograms of
\ P(M) and P(E) in Fig.1 (b) and (d). The shape of distribution P(E) shows a
double peak around the transition temperature corresponding to the
coexistence of the ordered and disordered phases. In the distribution of the
magnetization P(M), the middle peak indicate to the value of disordered
phase while the other peaks indicate the order phase. The histograms of P(M)
and P(E) around the transition temperature is given in Fig.2(a) and (b) for
a chosen simulation, respectively. Histograms of P(M) and P(E) have a single
peak above the transition temperature. As the temperature is close to the
transition value the histogram of P(E) has a double peak while P(M) has
three peaks indicate the coexistence region. With the decreasing
temperature, the histogram of P(E) converges to a single peak and the middle
peak in the histogram of P(M) is depressed while other peaks are enhanced
corresponding to order region. The behavior of P(M) and P(E) around the
transition region are in good agreement with the characteristic behavior of
the first order phase transition $^{(10,12,13,19)}$.

In Fig.3 (a) and (b) the temperature dependence of the order parameter, the
magnetic susceptibility, the energy and the specific heat are shown at
different lattice sizes for a selected simulation which shows a double peak
in the histogram of P(E) and three peaks in the P(M). The results are seen
in Fig.3 show that the energy and the order parameter have discontinuity at
transition region and the peaks of the susceptibility and the specific heat
increase with lattice size as expected in the first order phase transition.

The finite-size effects of the \ specific heat and the susceptibility at the
transition temperature are given by

\begin{equation}
C_{\max }\varpropto L^{d}
\end{equation}

\begin{equation}
\chi _{\max }\varpropto L^{d}
\end{equation}

The magnetic susceptibility maxima $\chi _{max}(L)$, the specific heat
maxima $C_{max}(L)$ are obtained from the average over the maximum of
magnetic susceptibility and specific heat at each lattice size for chosen
simulations. However, the transition temperature $T_{t}(L)$ for each lattice
is estimated from the location of the magnetic susceptibility and the
specific heat maxima for chosen simulations. The infinite lattice transition
temperature are estimated from the extrapolation of susceptibility $%
T_{t}^{\chi }(L)$ and specific heat maxima $T_{t}^{C}(L)$ for various
lattice sizes. The estimated infinite lattice transition temperature values
are $T_{t}^{\chi }(\infty )=1.29\pm 0.02$, $T_{t}^{C}(\infty )=1.28\pm 0.03$
from $\chi _{max}$ and $C_{max}$, respectively. The logarithm of the
specific heat and the susceptibility maxima as a function of the logarithm $%
L $ is shown in Fig.4(a) and (b). The data lies on a single curve and the
slope gives 3.01 and 3.07 from the specific heat maxima $C_{max}(L)$ and the
susceptibility maxima $\chi _{max}(L)$, respectively. These estimated values
are in good agreement with the system dimension ($d=3$). This result show
that, the magnetic susceptibility and the specific heat maxima increase with
the system volume ($L^{d}$) as a characteristic property of the first order
phase transition$^{(12-18)}$.

\textbf{4. Conclusion}

The three dimensional Blume-Capel model is simulated using cooling
algorithms on a cellular automaton. The simulations are done about $10$
times with different initial configurations at each lattice size and some of
this simulations having double and three peaks in the histograms of P(E) and
P(M) are chosen. To identify the phase transition of the model, the analysis
of the first order phase transition effects and the histogram technique are
used. The temperature variations of the order parameter and the energy show
the discontinuity at the transition temperature while the histogram of \
P(E) and P(M) have a double and three peak structures. Furthermore, the
magnetic susceptibility and the specific heat maxima increase with the
system volume ($L^{d}$\bigskip ) as expected in the first order phase
transition.

\textbf{Acknowledgements}

This work is supported by a grant from Gazi University (BAP:05/2003-07).

\textbf{References}

[1] M.B. Blume, Phys. Rev. B 141 (1966) 517.

[2] H.W. Capel, Physica(Utrecht) 32 (1966) 966.

[3] D.M. Saul, M. Wortis and D.Stauffer, Phys. Rev. B 9 (1974) 4964.

[4] J.G. Brankov, J. Przystawa and E. Pravecki, J. Phys. C 5 (1972) 3384.

[5] W.M. Ng and J.H.Barry, Phys. Rev. B 17 (1978) 3675.

[6] A.F. Siqueira and I.P. Fittipaldi, Physica A 138 (1986) 592.

[7] A. Du, Y.Q. Y\"{u} and H.J. Liu, Physica A 320 (2003) 387.

[8] J.W. Tucker, J.Phys.:Condens. Matter I, (1989) 485.

[9] B. Kutlu, A. \"{O}zkan, N. Sefero\u{g}lu, A. Solak and B. Binal, Int. J.
Mod. Phys.C 16 ( 2005) (in press).

[10] O.G. Mouritsen, "Computer Studies of Phase Transitions and Critical
Phenomena", New York Tokyo, 1984.

[11] V.Privman, "Finite Size Scalin and Numerical Simulation of Statistical
Systems", Singapore, 1990.

[12] M.S.S.Challa, D.P.Landau and K.Binder, Phys. Rev. B 34 (1986) 1841.

[13] K.Binder and D.P.Landau, Phys. Rev. B 30 (1984) 1477.

[14] S.Chen, A.M.Ferrenberg and D.P.Landau, Phys. Rev. Lett. 69 (1992) 1213.

[15] J. Lee and M.Kosterlitz, Phys. Rev. B 43 (1991) 3265.

[16] M.E. Fisher and A.N.Berker, Phys. Rev. B 26 (1982) 2507.

[17] V. Privman and M.E.Fisher, \ J. of Stat. Phys. 33 (1983) 385.

[18] J.L. Cardy and M.P. Nightingale, Phys. Rev. B 27 (1983) 4256.

[19] I. Puha and H.T. Diep, J. Magn. Magn. Mater. 224 (2001) 85.

\textbf{Figure Captions}

Fig.1. The temperature dependence of (a) the order parameters M and Q, (b)
the histogram of P(M)\ ,(c) the energy, (d) the histogram of P(E) for\ a
chosen simulation on L=18.

Fig.2. The histogram (a) P(M) and (b) P(E) for several temperatures around
the transition temperature for choosen simulations on L =10 and L = 18.

Fig.3. The temperature dependence of (a) the order parameters, (b) magnetic
susceptibility, (c) the energy and (d) the specific heat on L =
8,10,12,14,16,18 and 20.

Fig.4. Log-log plot of (a) the specific heat maxima and (b) the magnetic
susceptibility maxima against linear dimension L.

\bigskip \FRAME{ftbpF}{6.4359in}{6.8528in}{0pt}{}{}{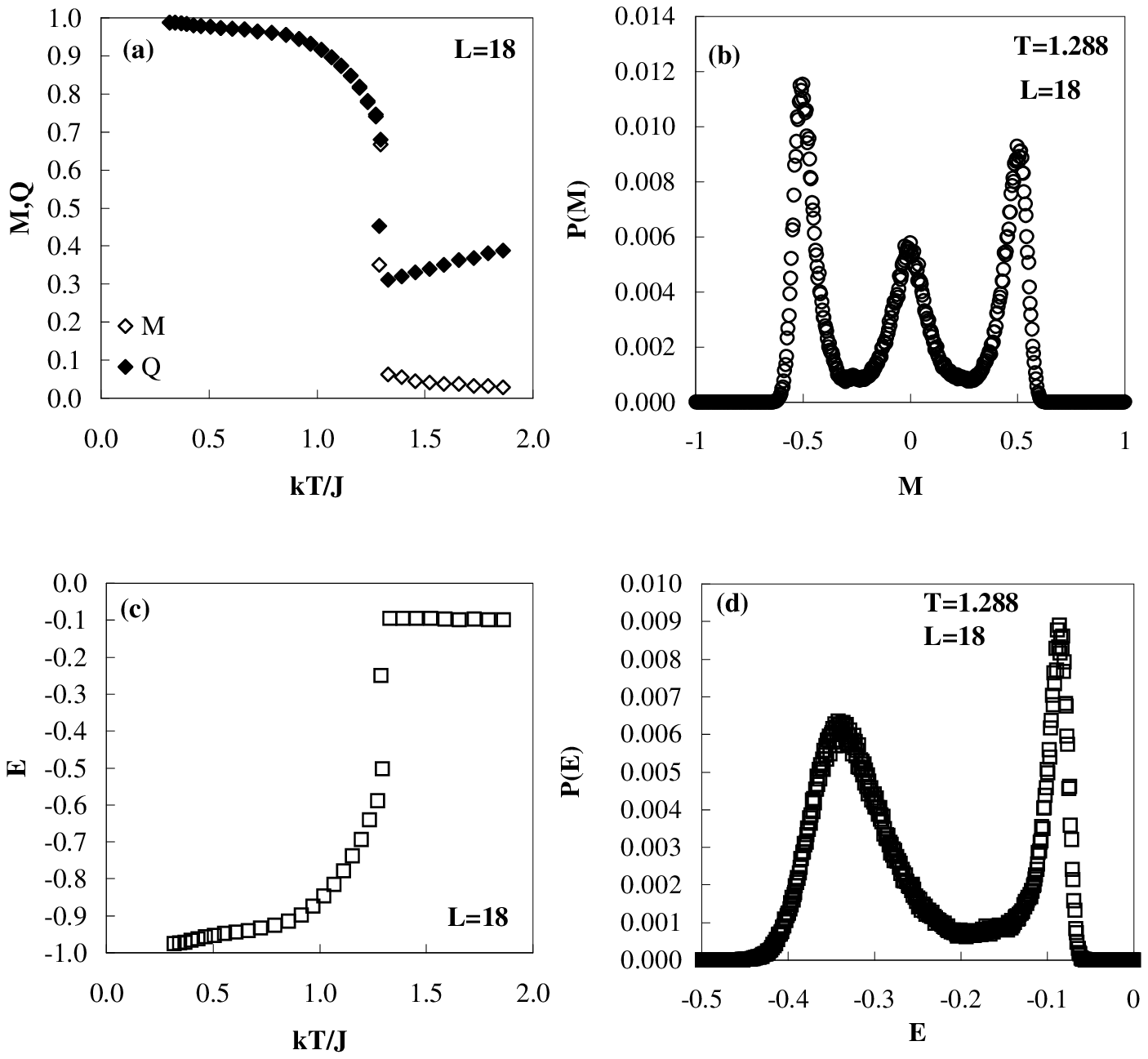}{\special%
{language "Scientific Word";type "GRAPHIC";display "USEDEF";valid_file
"F";width 6.4359in;height 6.8528in;depth 0pt;original-width
5.6654in;original-height 9.5372in;cropleft "0";croptop "1";cropright
"1";cropbottom "0";filename 'fig1.eps';file-properties "XNPEU";}}

\bigskip

\bigskip

\bigskip \FRAME{ftbpF}{6.0044in}{7.9641in}{0pt}{}{}{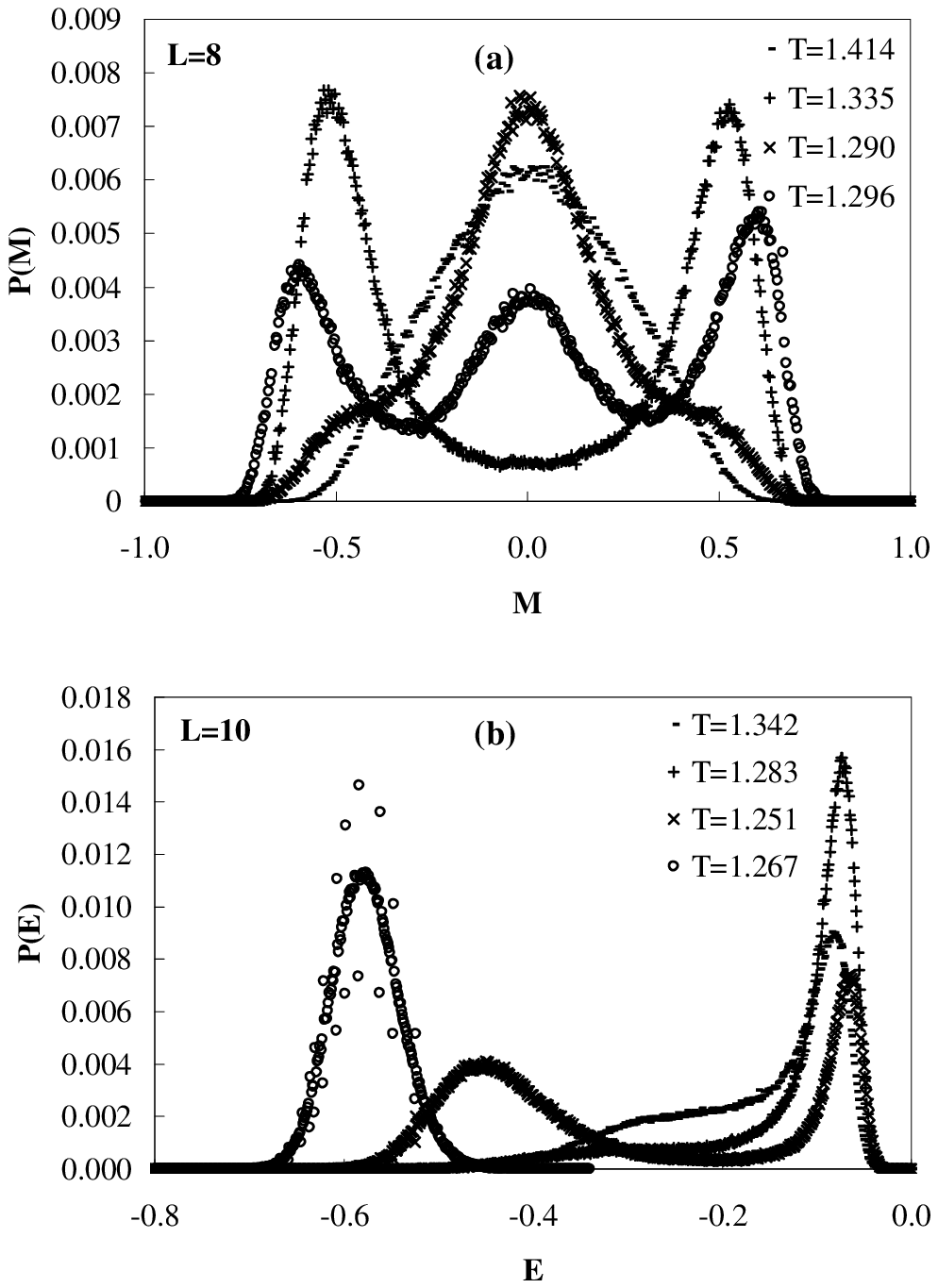}{\special%
{language "Scientific Word";type "GRAPHIC";display "USEDEF";valid_file
"F";width 6.0044in;height 7.9641in;depth 0pt;original-width
4.0171in;original-height 5.5711in;cropleft "0";croptop "1";cropright
"1";cropbottom "0";filename 'fig2.eps';file-properties "XNPEU";}}

\bigskip

\FRAME{ftbpF}{6.7706in}{7.5386in}{0pt}{}{}{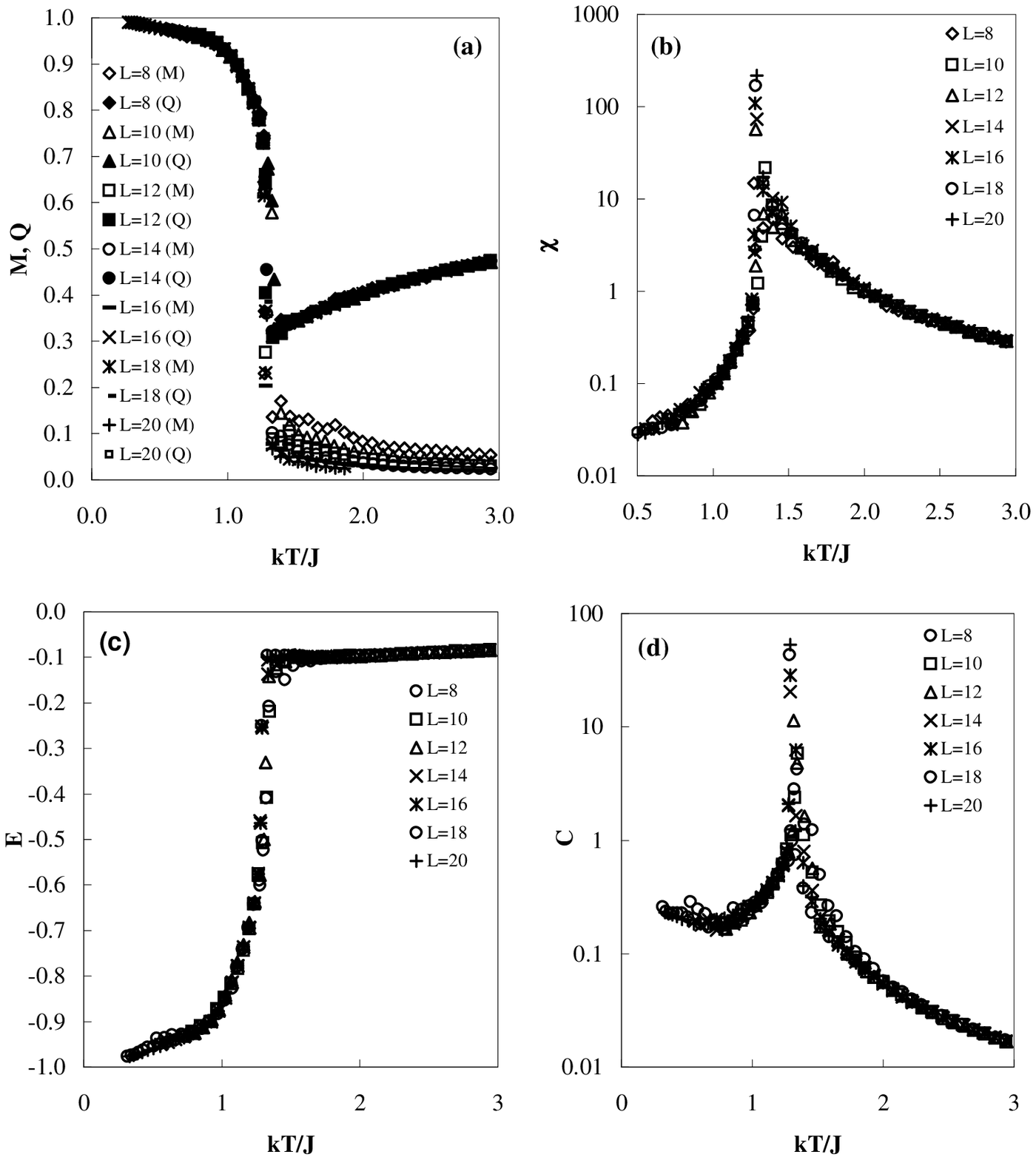}{\special{language
"Scientific Word";type "GRAPHIC";display "USEDEF";valid_file "F";width
6.7706in;height 7.5386in;depth 0pt;original-width 7.7721in;original-height
11.1855in;cropleft "0";croptop "1";cropright "1";cropbottom "0";filename
'fig3.eps';file-properties "XNPEU";}}

\bigskip \FRAME{ftbpF}{5.5106in}{7.9156in}{0pt}{}{}{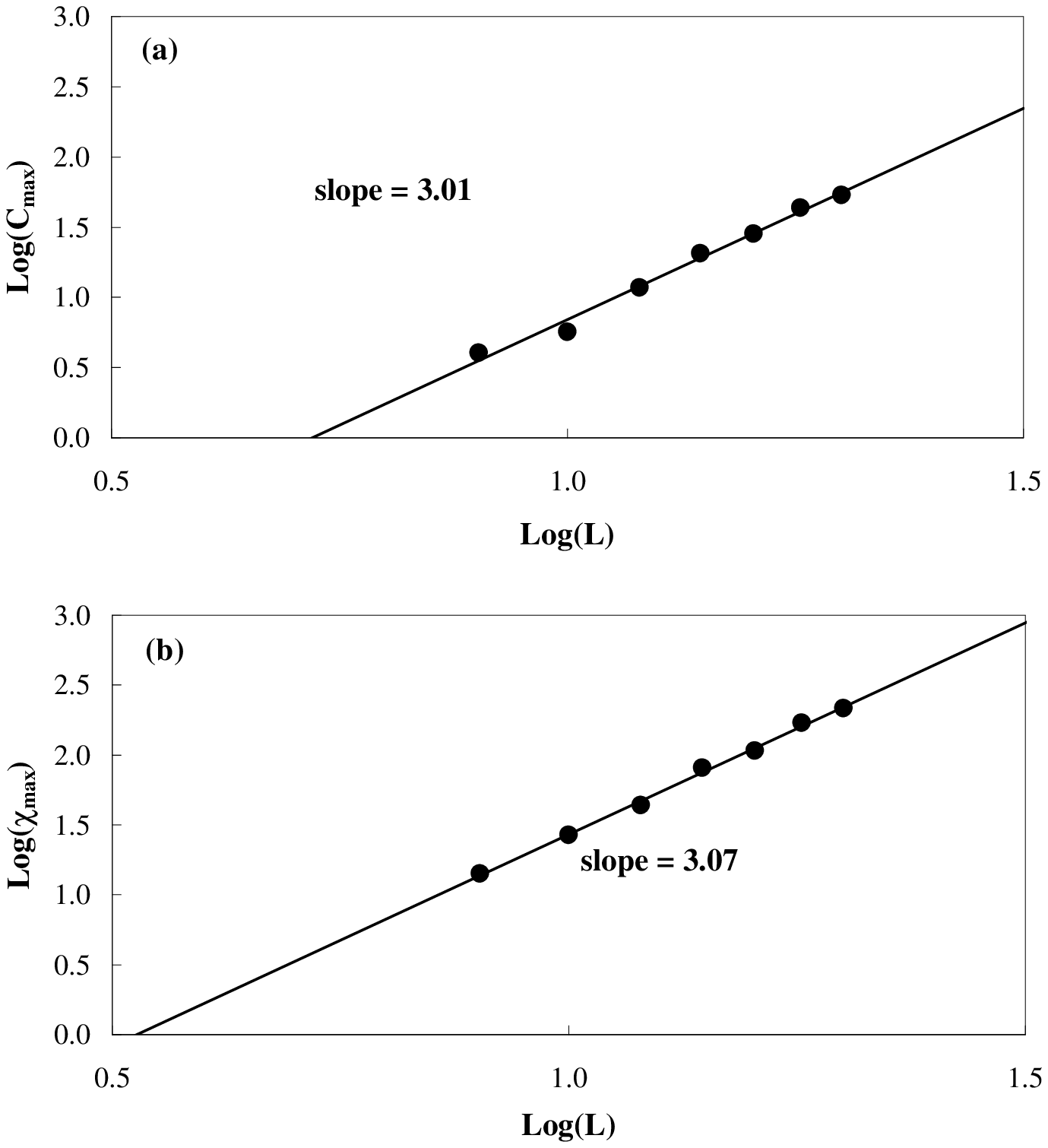}{\special%
{language "Scientific Word";type "GRAPHIC";display "USEDEF";valid_file
"F";width 5.5106in;height 7.9156in;depth 0pt;original-width
7.7721in;original-height 11.1855in;cropleft "0";croptop "1";cropright
"1";cropbottom "0";filename 'fig4.eps';file-properties "XNPEU";}}

\bigskip

\end{document}